\documentclass[fleqn]{elsarticle}  
\usepackage{color}		
\usepackage{epsfig}
\usepackage{natbib}
\usepackage{amsmath}
\usepackage{amssymb}
\usepackage{graphicx}
\begin{document}
\title{Nonlinear chaos in temperature time series:\\Part I:\\ Case studies}
\author{Yaron ~Rosenstein}
\ead{yrosenstein@gmail.com}
\author{Gal ~Zahavi}
\ead{galz@ie.technion.ac.il}

\begin{abstract}
In this work we present 3 case studies of local temperature time series obtained from stations in Europe and Israel.
The nonlinear nature of the series is presented along with model based forecasting.
Data is nonlinearly filtered using high dimensional projection and analysis is performed on the filtered data.
A lorenz type model of 3 first order ODEs is then fitted.
Forecasts are shown for periods of 100 days ahead, outperforming any existing forecast method known today.
While other models fail at forecasting periods above 11 days, ours shows remarkable stability 100 days ahead.

Thus finally a local dynamical system if found for local temperature forecasting not requiring solution of Navier-Stokes equations.
Thus saving computational costs.

\end{abstract}
\maketitle

\section{Introduction}
Weather forecasting is crucial for government, military, industrial and agricultural applications.
As well as an open scientific question.
\newline
Global numerical weather forecasting pioneered by ~\cite{Richardson} today is carried out using a multitude of methods, each having it's advantages and disadvantages.
The foremost method consists of the solution of the global flow and energy equations of the atmosphere (Navier-Stokes).
Due to the inherent complexity of these equations, simplifications are used.
These include hydrostatic, geostrophic and quasi-geostrophic approximations.
The domain is then divided using a grid and equations are solved within the grid and advanced in time, using finite elements 
or finite difference methods.
Newer methods include spectral methods, which show exponential convergence for smooth problems ~\cite{Simmons}. 

Manipulating the vast datasets and performing the complex calculations necessary to modern numerical weather prediction
requires some of the most powerful supercomputers in the world. 
Even with the increasing power of supercomputers, the forecast skill of numerical weather models extends to about
only six days. Factors affecting the accuracy of numerical predictions include the density and quality 
of observations used as input to the forecasts, along with deficiencies in the numerical models themselves. 
Although post-processing techniques such as model output statistics (MOS) have been developed to improve the handling 
of errors in numerical predictions, a more fundamental problem lies in the chaotic nature of the 
partial differential equations used to simulate the atmosphere. 
It is impossible to solve these equations exactly, and small errors grow with time (doubling about every five days). 
In addition, the partial differential equations used in the model need to be supplemented with 
parameterizations for solar radiation, moist processes (clouds and precipitation), heat exchange, soil, vegetation, 
surface water, and the effects of terrain. 
In an effort to quantify the large amount of inherent uncertainty remaining in numerical predictions, 
ensemble forecasts ~\cite{Gneiting} have been used since the 1990s to help gauge the confidence in the forecast, 
and to obtain useful results farther into the future than otherwise possible. 
This approach analyzes multiple forecasts created with an individual forecast model or multiple models.
Monte Carlo simulations are then carried out on the forecasts to obtain the statistical distribution of error of the forecast.
The limitations of ensemble methods are inherent in the global modeling approach.

Other methods for weather forecasting include the ``downscaling'' of global climate models (GCMs) ~\cite{Wilby} to localized regions
using statistical relations between \textit{predictors} and \textit{predictands}. Predictors are taken from the GCM model output such as temperature, pressure at different altitudes and locations.
Predictands are the local variables to be simulated, these include temperature, pressure, wind and precipitation.
These methods are mainly used for the study of impact of global warming on surface variables.
~\cite{Palutikof} have shown that these models deviate significantly from observed data.
Statistical methods of weather forecasting from time series are known from the works of ~\cite{Alaton} and 
~\cite{Campbell}.
In these methods the series is decomposed into cyclical, trend and error terms.
The cyclical component is approximated by a Fourier decomposition, trend with linear term and error is approximated using 
GARCH model.
These models however don't teach us anything about the dynamics of the system.
Further more Garch models explode at prediction ranges larger than 11 days.

In this work we present for the first time the existence of a deterministic ODE system describing the evolution of temperature 
from daily mean temperature time series data.
Section 2 describes the embeding procedure, section 3 describes the model fitting and shows results for forecasting 100 days forward.

\section{Mathematical model}
\subsection{Nonlinear filtration of time series}

We assume that the time series are generated by a dynamical system of the form:
\begin{equation}
\label{dynsys}
 \frac{dx_i}{dt}=f_i\left(\{x_j\}_{j=1}^m\right)+\xi_i,\; i=1, 2, 3, ..., m
\end{equation}
Where $\xi_i$ is the inherent noise of the system which is assumed to be i.i.d. (independent, identically distributed).

And the observable has measurement noise:
\begin{equation}
y(t)=\eta\left(\{x_j\}_{j=1}^m\right)+\xi
\end{equation}

In order to obtain the dynamical system, data has to be filtered first.
We use high-dimensional projection and singular value decomposition (SVD) ~\cite{Sauer3} to obtain the attractor from the principal directions.
I.e. noise is projected to high dimensional manifold (dimension 12).
Singular value decomposition is carried out on the projection to obtain principal directions of attractor.

This is then projected back to time series as clean signal, filter takes about 250 iterations to converge to clean signal.

Numerical analysis was carried out using the TISEAN package ~\cite{Hegger}.
TISEAN is a nonlinear time series analysis package developed on the principles of ~\cite{Sauer1}, ~\cite{Procaccia} and ~\cite{Kantz}.

\begin{figure}[htbf]
\begin{center}
\includegraphics[scale=0.7]{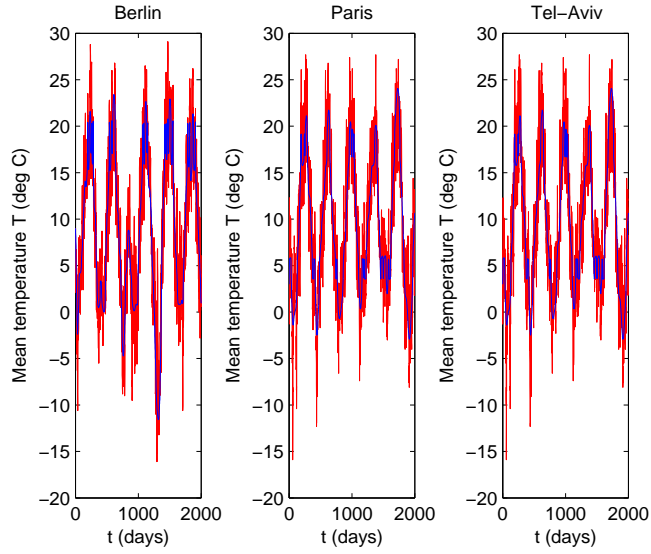}
\end{center}
\caption{Filtered(blue) vs. raw data(red) of 3 stations, Berlin, Paris, Tel-Aviv, 2000 points}
\label{rawfiltered}
\end{figure}

\begin{figure}[htbf]
\begin{center}
\includegraphics[scale=0.7]{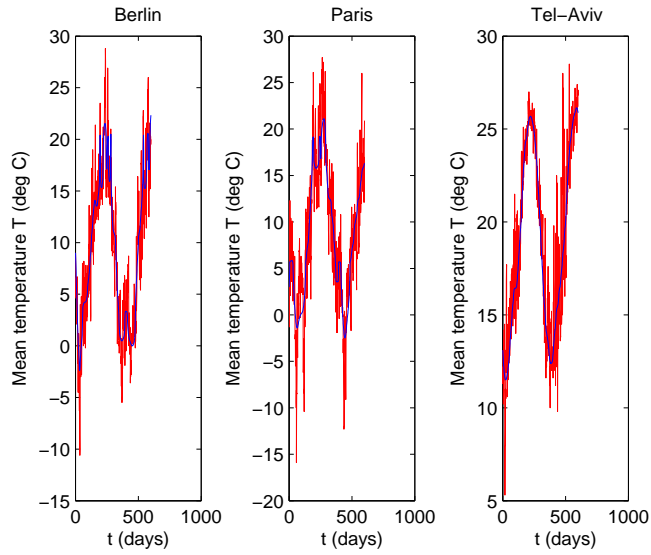}
\end{center}
\caption{Filtered(blue) vs. raw data(red) of 3 stations, Berlin, Paris, Tel-Aviv, zoomed to 600 points}
\label{rawfilteredzoomed}
\end{figure}

\begin{figure}[htbf]
\begin{center}
\includegraphics[scale=0.7]{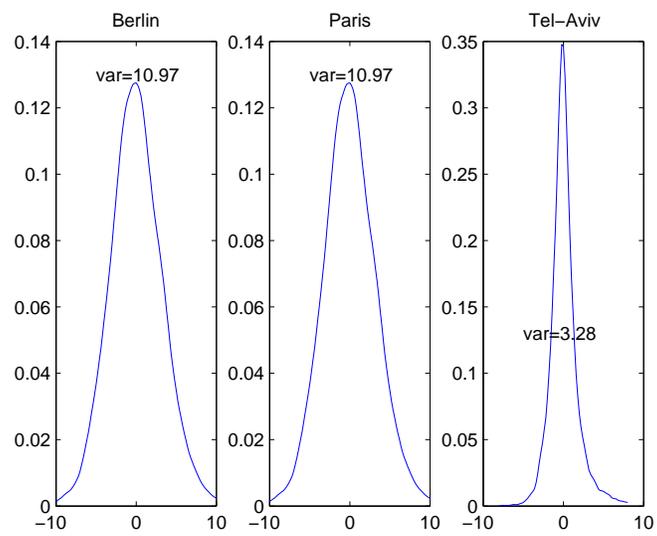}
\end{center}
\caption{Histograms of residuals of Berlin, Paris, Tel-Aviv, variances are shown in the figure}
\label{rawfilteredhist}
\end{figure}

We sampled mean daily temperature from a few stations. 
Data was obtained from European weather center http://www.knmi.nl.
Each time series consists of 15000 terms.
Figs. ~\ref{rawfiltered}, ~\ref{rawfilteredzoomed} and ~\ref{rawfilteredhist} show an excellent fitting of the filtered dynamical system to raw data.
The first one depicting the original data vs. filtered on 2000 points for Berlin, Paris and Tel-Aviv respectively.
The second figure shows the first 600 points of fig. ~\ref{rawfiltered}, remarkable fit is obtained.
The third, fig. ~\ref{rawfilteredhist} shows the distribution of residuals:
\begin{equation}
\label{resi1}
 r(t)=y_f(t)-y(t)
\end{equation}
Variances being 10.97, 10.97, 3.28 for Berlin, Paris and Tel-Aviv 
respectively.

\clearpage
\subsection{Embedding results}
We are given the filtered time series $x(t_i)$ from previous section which is assumed to be 
generated 
by (~\ref{dynsys}), possesing an attractor $A$ with box dimension $d_A$.
\newline
Whitney's embedding theorem ~\cite{Whitney} states that any smooth $n$ dimensional manifold can be embedded in $2n+1$ Euclidean space. I.e. no two points on the manifold map to the same point in $R^{2n+1}$
\newline
Takens theorem ~\cite{Takens} extends Whitney's theorem such that an attactor $A$ with box dimension $d_A$ can be embedded in $k$ dimensional Euclidean space provided $k>2d_A$.
\newline
I.e. there exists a diffeomorphism $\phi$ from $A$ into $R^k$.
The delay theorem states that using the delay vector:
\[
\left(x(t),x(t-\tau),x(t-2\tau),...,x(t-(k-1)\tau)\right)
\]
where $\tau$ is the delay, reconstructs the system in $R^k$ if the dimension of system $2d_A\leq k$.
Hence 2 parameters govern the reconstruction of attractor, namely $\tau,m$ the delay and dimension of space.
We use autocorrelation function $R_x(\tau)$ to compute the delay which matches the decorrelation time \cite{Sauer1}, i.e. when $R_x(\tau)\leq \frac{1}{e}$.
\begin{equation}
R_x(\tau)=\frac{E\left[(x(t)-\mu)(x(t+\tau)-\mu)\right]}{\sigma^2}
\end{equation}
It is evident that $-1\leq R_x(\tau)\leq 1$.
$R_x$ can also be defined using the convolution theorem:
\begin{equation}
R_x(\tau)=\frac{1}{2\pi}\int_{-\infty}^{\infty}{S(\omega)e^{-i \omega \tau}d\omega}
\end{equation}
Where:
\begin{align*}
S(\omega)&={\mathcal F}(x)^*{\mathcal F}\\
{\mathcal F}(x)&=\int_{-\infty}^{\infty}{x(t)e^{i \omega t}dt}
\end{align*}
The decorrelation time $\tau$ has physical significance, it is defined as the time at which the phase of the wave has wandered significantly and the wave is no longer coherent.
Formally it is defined as the inverse of the spectral width of the signal.
\begin{align}
\Delta \nu=\frac{\left(\int_{0}^{\infty}{S(\nu)d\nu}\right)^2}{\int_{0}^{\infty}{S^2(\nu)d\nu}}\\
\tau=\frac{1}{\Delta \nu}
\end{align}

For deterministic monochromatic waves the decorrelation time is infinite.
For white Grassbergernoise:
\begin{equation}
R_x(\tau)=\delta(0)
\end{equation}
Decorrelation is immediate.

We use false nearest neighbor analyis ~\cite{Kennel} to determine the proper embedding dimension.
The idea is quite intuitive. Suppose the minimal embedding dimension for a given time series i s $m_0$. This means that in a $m_0$-dimensional delay space the reconstructed attractor is a one-to-one image of the attractor in the original phase space. Especially, the topological properties are preserved. Thus the neighbors of a given point are mapped onto neighbors in the delay space. Due to the assumed smoothness of the dynamics, neighborhoods of the points are mapped onto neighborhoods again. Of course the shape and the diameter of the neighborhoods is changed according to the Lyapunov exponents. But suppose now you embed in an $m$-dimensional space with $m<m_0$. Due to this projection the topological structure is no longer preserved. Points are projected into neighborhoods of other points to which they wouldn't belong in higher dimensions. These points are called false neighbors. If now the dynamics is applied, these false neighbors are not typically mapped into the image of the neighborhood, but somewhere else, so that the average ``diameter'' becomes quite large. 
For each point in the series we compute the nearest points and pick the one with the minimal Euclidean distance and compute the distance in the next iteration.
\begin{equation}
R_i=\frac{\|\mathbf{x}_{i+1}-\mathbf{x}_{j+1}\|}{\|\mathbf{x}_{i}-\mathbf{x}_{j}\|}
\end{equation}
If $R_i$ exceeds a certain heuristic then $x_j$ is a false neighbor.

When the fraction of false nearest neighbors shows first local minimum the dimension $m$ is determined.

The correlation dimension is computed using ~\cite{Procaccia}.
The correlation integral:
\begin{equation}
C(r)=\int{d\mu(\mathbf{x})\int{\Theta(r-\|\mathbf{x}-\mathbf{y}|)d\mu(\mathbf{y})}}
\end{equation}
Where $\mu$ is the probability measure of the set of points and $\Theta$ is the Heaviside step function.
Basically $C(r)$ counts points of distance less than $r$.
It is assumed in~\cite{Procaccia} that:
\begin{equation}
C(r)\sim r^D
\end{equation}
Thus the correlation dimension is defined as:
\begin{equation}
D=\lim_{r\rightarrow 0}{\frac{log(C(r))}{log(r)}}
\end{equation}

\begin{figure}
\begin{center}
\includegraphics[scale=0.7]{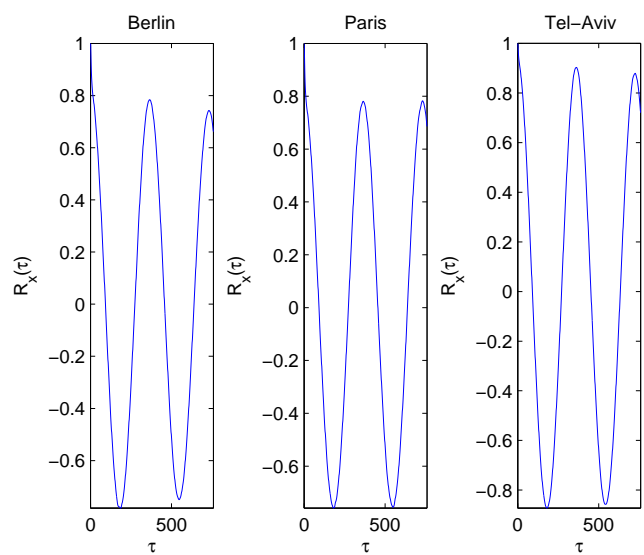}
\end{center}
\caption{Plot of autocorrelation functions for Berlin, Paris and Tel-Aviv respectively}
\label{autocor}
\end{figure}

Fig. ~\ref{autocor} shows the autocorrelation function $R_x(\tau)$ as function of $\tau$.
Decorrelation delay is $\tau=65$.
Further more, local maxima at $\tau=365n$ where $n$ is an integer, indicate annual periodicity of time series.
It is notable that all stations display same characteristics.

\clearpage

\begin{figure}
\begin{center}
\includegraphics[scale=0.7]{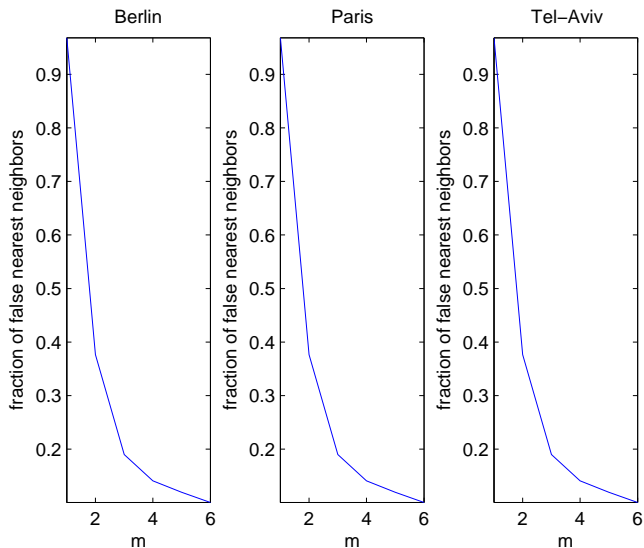}
\end{center}
\caption{False nearest neighbors fraction, Berlin, Paris and Tel-Aviv stations}
\label{fnn}
\end{figure}

We also performed false nearest neighbor analysis on the readings.
Since the data is assumed noisy we look at the first minimum.
Fig. ~\ref{fnn} show the dimension in all stations to be $m=3$.

Next we performed the actual delay map and computed the positive Lyapunov exponents and Grassberger-Procaccia ~\cite{Procaccia} correlation dimenions which are the topological invariants of attractors.
\begin{figure}
\begin{center}
\includegraphics[scale=0.7]{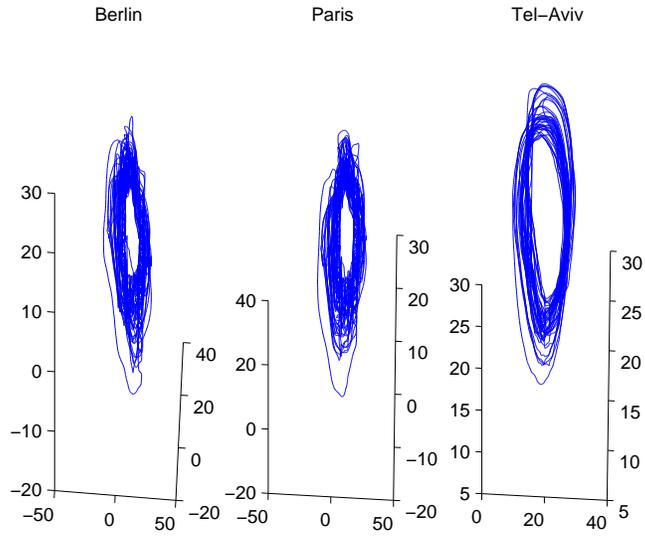}
\end{center}
\caption{Reconstructed attractor, Berlin, Paris and Tel-Aviv stations}
\label{att}
\end{figure}

\begin{figure}
\begin{center}
\includegraphics[scale=0.7]{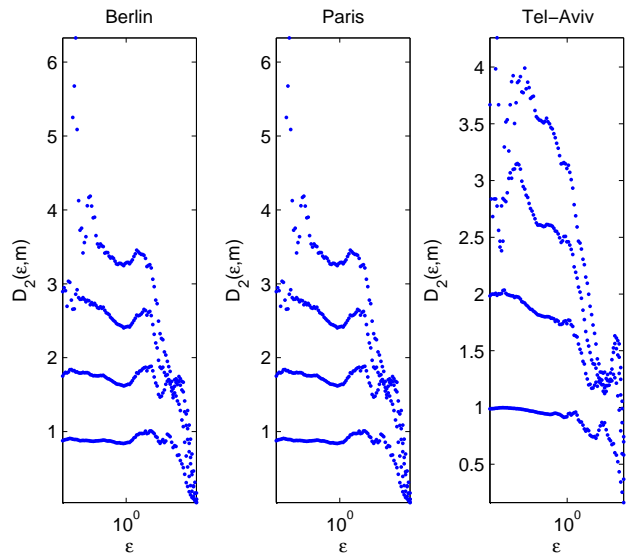}
\end{center}
\caption{Correlation sums for Berlin, Paris and Tel-Aviv stations}
\label{corrsums}
\end{figure}
\begin{figure}
\begin{center}
\includegraphics[scale=0.7]{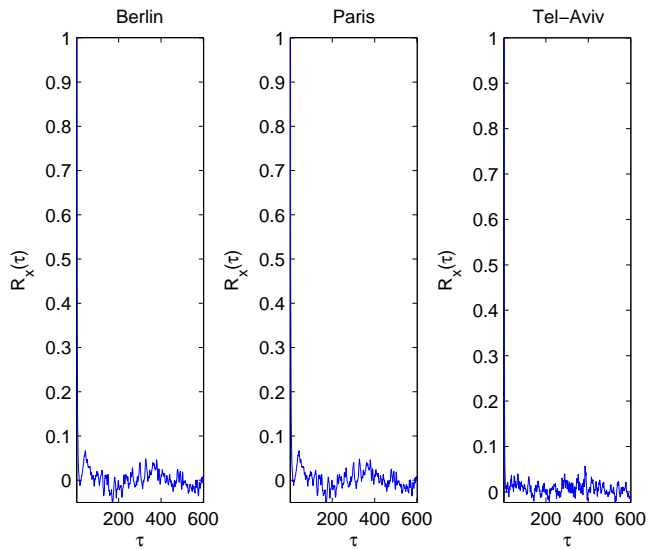}
\end{center}
\caption{Autocorrelation functions of $r(t)$, Berlin, Paris and Tel-Aviv stations}
\label{rescor}
\end{figure}

Leading Lyapunov exponents:
Table ~\ref{lyap} shows the leading Lyapunov exponent for the stations.
\begin{table}[htbp]
\label{lyap}
\caption{Positive Lyapunov exponents for Tel-Aviv, Paris and Berlin}
\begin{tabular}{|p{5cm}|p{5cm}|p{5cm}|} \hline
\textbf{Berlin} & \textbf{Paris} & \textbf{Tel-Aviv}\\ \hline
0.246 & 0.250 & 0.062\\ \hline
0.242 & 0.243 & 0.060\\ \hline
\end{tabular}
\end{table}
\begin{table}[htbp]
\label{proc}
\caption{Grassberger-Procaccia correlations for Tel-Aviv, Paris and Berlin}
\begin{tabular}{|p{5cm}|p{5cm}|p{5cm}|} \hline
\textbf{Tel-Aviv} & \textbf{Paris} & \textbf{Berlin}\\ \hline
1.6 & 1.7 & 1.3\\ \hline
\end{tabular}
\end{table}
Figs. ~\ref{autocor} to ~\ref{corrsums} and tables ~\ref{lyap} and ~\ref{proc} show that the same dynamical system 
controls the local temperature at 3 different stations on earth.
We have also computed the autocorrelation function of the residual noise $r(t)$, equation (~\ref{resi1}).
Fig. ~\ref{rescor} shows that the autocorrelation function for all 3 stations is that of white noise.
This establishes our claim at equation (~\ref{dynsys}).

\section{Model based prediction}
\subsection{Fitting of set of first order ODEs}
We assume our reconstructed dynamical system obeys bilinear (generalized Lorenz) + 
trend + 
365 days period forcing terms:
\begin{equation}
\frac{dx_i}{dt}=f_i(\{x_j\}_{j=1}^3,\alpha),\; i=1, 2, 3
\end{equation}
\begin{equation}
f_i(\{x_j\}_{j=1}^3,\alpha)=\alpha_{i00}+\alpha_{ij0}x_j+\alpha_{ijk}x_j x_k+\beta_i\cos{\frac{2\pi \left(t\bmod 
365\right)}{365}}
\end{equation}
Where summation convention is applied.

We are given the phase reconstructed data from previous section fig. ~\ref{att}.

We need to match the $\mathbf{\alpha}$ parameters tensor and the 
given data.
Let us vectorize the tensor $\alpha$.
\begin{equation}
 \alpha_{l=k+3j+9i}^*=\alpha_{ijk}
\end{equation}

\begin{figure}
\begin{center}
\includegraphics[scale=0.7]{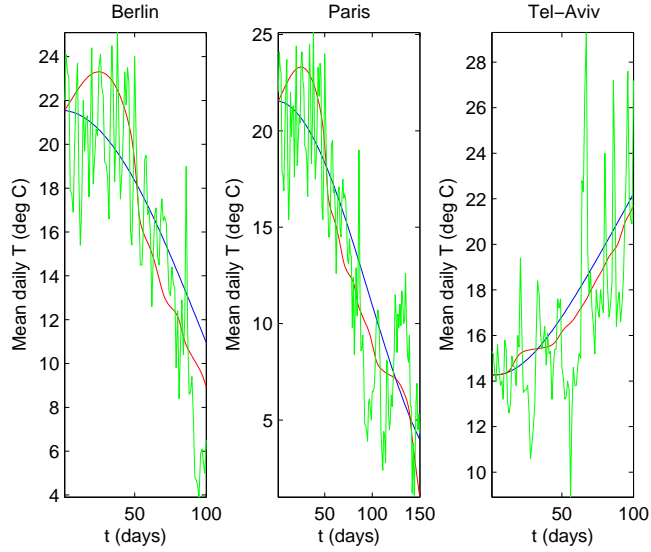}
\end{center}
\caption{Measured (green) vs. filtered (red) and predicted solution of odes (blue) for Berlin, Paris and Tel-Aviv}
\label{fit}
\end{figure}

\begin{figure}
\begin{center}
\includegraphics[scale=0.7]{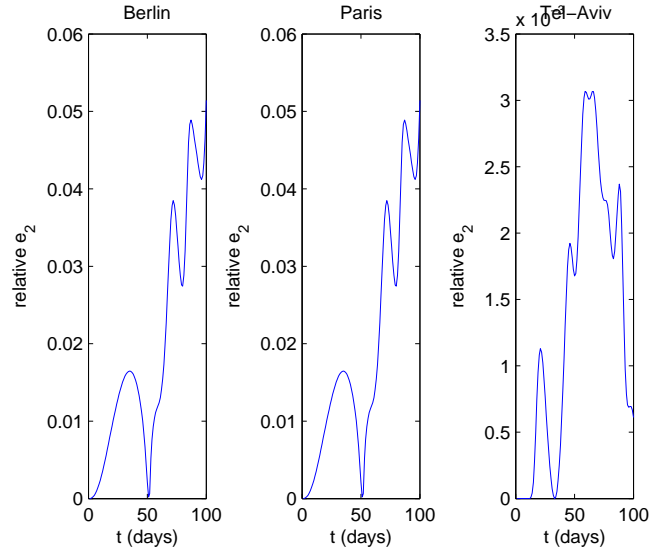}
\end{center}
\caption{Relative squared error for fit vs. filtered data for Berlin, Paris and Tel-Aviv}
\label{err}
\end{figure}

We use the method employed by ~\cite{Baker} and define the cost function:
\begin{equation}
H(\{x_i\}_{i=1}^3,\mathbf{\alpha^*})=\sum_{i=1}^n{\sum_{j=1}^3{\left(x_j'(t_i)-f_j(\mathbf{\alpha^*})\right)^2}}
\end{equation}
Gauss-Newton method ~\cite{Baker} is used to minimize the cost function $H(x,y,z,\mathbf{\alpha^*})$.
Assume an initial parameter vector $\alpha_0^*$ and
use the Taylor expansion of $H(x,y,z,\alpha^*)$ with respect to $\alpha^*(i)$ is:
\begin{equation}
\begin{split}
 H(\{x_i\}_{i=1}^3,\alpha^*)=H(\{x_i\}_{i=1}^3,\alpha_0^*)+\sum_{j=1}^m{\frac{\partial H}{\partial \alpha^*(j)}(\alpha^*(j)-\alpha_0^*(j))}+\\
\frac{1}{2}\sum_{j=1}^m{\sum_{k=1}^m{(2-\delta_{jk})\frac{\partial^2 H}{\partial \alpha^*(j)\partial\alpha^*(k)}(\alpha^*(k)-\alpha_0^*(k))(\alpha^*(j)-\alpha_0^*(j))}}+O(\delta\alpha^3)
\end{split}
\end{equation}
Next, we solve $m$ equations:
\begin{equation}
\frac{\partial H}{\partial \alpha^*(j)}=2\sum_{i=1}^n{\sum_{k=1}^3{\left(x_k'(t_i)-f_k(\mathbf{\alpha^*})\right)\frac{\partial f_k}{\partial \alpha^*(j)}}}=0
\end{equation}
The time derivatives are taken using difference scheme:
\begin{equation}
 x_a'(t_j)=\frac{-x_a(t_{j+2})+8x_a(t_{j+1})-8x_a(t_{j-1})+x_a(t_{j+1})}{12\Delta t}
\end{equation}
Performing the minimization procedure on learning set of 14000 days we obtain a solution for $\alpha$ .
System is then integrated forward in time, system is stiff (positive Lyapunov exponents (table ~\ref{lyap})) and solution explodes after 150 days ahead.
ODE fit is shown for Berlin, Paris and Tel-Aviv in figs. ~\ref{fit} and ~\ref{err}, 100 days ahead.
\newline
Fig. ~\ref{fit} shows the fit of 
solution of odes to reconstructed filtered data and raw data.
The trajectory fits well to the filtered data and passes in midline of oscillations of raw data.
Fig. ~\ref{err} shows the relative squared error of fitted vs. filtered data for prediction 100 days ahead.

\newpage
\section{Concluding remarks}
We have established in this work the existence of a deterministic dynamical system governing the evolution of local temperatures.
This dynamical system yields a very good fit to the nonlinearly filtered data.
The residual noise (difference between dynamical system and raw data) is bounded.
We have also shown using autocorrelation and distribution fittng that residual noise is Gaussian.
Further work is required to improve fitting of dynamical system to the filtered data.
Further work is also required to derive a complete stochastic differential equation with a mean reverting process.

\bibliographystyle{plain}

\end{document}